\documentstyle[12pt]{article}
\def\ga{\alpha} \def\gb{\beta} 

\def\gd{\delta}  \def\ge{\epsilon} 
\def\gw{\omega}  
       
\def\gm{\mu}   \def\gn{\nu}  \def\gs{\sigma} \def\gS{\Sigma} 
 \def\gl{\lambda}   
\def\n{\noindent} 
\def\np{Nucl. Phys. {\bf B}} 
\def\pl{Phys. Lett. {\bf B}}

\def\prd{Phys. Rev. {\bf D}}

\def\be{\begin{equation}}  \def\ee{\end{equation}}   
\def\it{\item}
\def\CC{\kern 0.27em \vrule height1.45ex width0.03em depth0em \kern-0.30em%
\rm C}
\newcommand{\ignore}[1]{}

\begin{document}

\hoffset = -1truecm
\voffset = -2truecm

\hoffset = -1truecm
\voffset = -2truecm

\title
{\bf 
\thanks{This work was partially supported by CONICET, Argentina} 
}  

\vskip 2cm

\author{ Adri\'an R. Lugo\thanks{Electronic address: 
lugo@dartagnan.fisica.unlp.edu.ar}\\
\normalsize Departamento de F\'\i  sica, Facultad de Ciencias Exactas\\
\normalsize Universidad Nacional de La Plata\\
\normalsize C.C. 67, (1900) La Plata, Argentina 
} 

\date{December 1996}
\vskip 2cm

\maketitle

\begin{abstract}
We compute the exact effective string vacuum backgrounds 
of the level $\;k = 81/19 \;\; SU(2,1)/U(1)\;$ coset model. 
A compact $\;SU(2)\;$ isometry present in this seven dimensional solution 
allows to interpreting it after compactification as a four dimensional 
non-abelian $SU(2)$ charged instanton with a singular submanifold 
and an $\; SO(3)\times U(1)\;$ isometry. 
The semiclassical backgrounds, solutions of the type II strings,  
present similar characteristics.
\end{abstract} 

\newpage

\n{\bf 1. Introductory remarks}
\bigskip

The search of interesting vacua representing the effective arena in which 
a string moves  has been during the last decade one of the most 
explored subjects in string theory [1].
The main reason behind this is to try to elucidate some natural mechanism 
of compactification from $26$ (or $10$ in the supersymmetric case) dimensions 
to the usual $4$, or in any case to have consistently defined a four 
dimensional theory, with the hope of obtaining string models compatible with 
well-known low energy physics [2]. 
In this context the  Kaluza-Klein (KK) mechanism naturelly arises. 
String solutions of this type comes in the form of exactly solvable 
two-dimensional sigma models known as gauged Wess-Zumino-Witten models 
(GWZWM's).
In Reference [3] we studied this mechanism in an abelian case.
In this letter we present a non trivial example of it based on the 
$SU(2,1)/U(1)$ coset model that gives rise to non abelian $SU(2)$ gauge 
fields. 
\bigskip

Let us start remembering some relevant facts about non abelian KK 
dimensional reduction [4].
Let us assume that our fields live on a $d+N$ dimensional manifold of the 
form $\; M\times \gS\;$ where the ``space-time" $M$ has dimension $d$ and 
the compact $\gS$ is $N$ dimensional, and let us restrict ourselves to 
product-like coordinate patches $\{ (z^M ) = (x^\gm, y^m ), \;\gm = 1, 
\dots , d\; ,\; m = 1,\dots ,N \}$.

Let $G_i$ be a group of diffeomorfisms of $\gS$ (and then, of 
$M\times\gS$) of dimension $d_i$ 
\begin{eqnarray}
x^\gm &\rightarrow& x'^\gm(x) = x^\gm \cr
y^m &\rightarrow& y'^m (y) = y^m + \ge^a\; u_a^m(y) + {\cal O}(\ge^2)
\end{eqnarray}
generated by the vector fields 
\begin{eqnarray}
u_a &=& u_a^m(y)\; \partial_m  \;\;\;\; , \;\;\; a = 1,\dots , d_i\cr
[u_a , u_b ] &=& f^c{}_{ab}\; u_c
\end{eqnarray}
where $f^c{}_{ab}$ are the structure constants of ${\cal G}_i $, 
$( X_a )^c{}_b = f^c{}_{ab} $ being the adjoint representation. 
We assume they are complete on $T_0^1(\gS |_y )$, the tangent space of 
$\gS$ at $y$ (and then $d_i \geq N$). 
Under this hypothesis, a $G_i$-invariant scalar field $D$ is just a field 
independent of the coordinates $(y^m) $.
Let $G_g$ be the group of diffeomorfisms of $\gS$ of 
dimension $d_g$ 
generated by the vector fields that are {\cal invariant} under $G_i$, i.e.
\footnote{
That they generate the Lie algebra of a group is easily showed.  
${\cal L}_{u_a}$ stands for the Lie derivative [5]. 
In usual notation the last equation in (3) reads 
$$
v_{\overline a}^m (y') =  \partial_n y'^m (y) \; v_{\overline a}^n ( y )
$$
with $y'$ given by (1).
}
\begin{eqnarray}
v_{\overline a}  &=& u_{\overline a}^m (y)\; \partial_m  \;\;\; ,\;\;  
\overline a = 1, \dots , d_g \cr
[v_{\overline a} , v_{\overline b} ] &=& 
{\overline f}^{\overline c}{}_{\overline a \overline b }\; v_{\overline c}\cr
{\cal L}_{u_a} ( v_{\overline b} ) &=& [u_a , v_{\overline b} ] = 0 
\end{eqnarray}
where ${\overline f}^{\overline c}{}_{\overline a\overline b }$ are the 
structure constants of ${\cal G}_g $.

\ignore{
With the help of the $v_{\overline a}$ we can introduce  $G_i$-invariant 
forms by 
\begin{eqnarray}
\ga^m &=& dy^m + A^{\overline a} \; v_{\overline a}^m\cr   
A^{\overline a} &=& dx^{\gm} \; A_\gm^{\overline a}(x) 
\end{eqnarray}
where $A^{\overline a}$ is invariant under $G_i$
\footnote{
In particular, $y$-independent, see below.
} 
\be
{\cal L}_{u_a} \; C^{\overline b} = 0
\ee 
and otherwise arbitrary. 
}
Let us consider now an arbitrary tensor field $t \in \tau_2^0 (\gS )$ 
\begin{eqnarray}
t &=& t^{M N}(z) \; dz^M \otimes dz^N \cr
&=& t_{\gm\gn} \; dx^\gm\otimes dx^\gn + t_{mn}\; dy^m\otimes dy^n 
+ t_{\gm m}\; dx^\gm\otimes dy^m + t_{m\gm} \; dy^m\otimes dx^\gm
\end{eqnarray}
If we now ask for $t$ to be $G_i$-invariant, then the set of equations
\be
{\cal L}_{u_a} ( t_{MN} ) = 0 
\ee
should be fullfilled; the solution of (5) leads to the following general 
form for $t$
\be 
t = t_{\gm\gn}(x)\; dx^\gm\otimes dx^\gn + 
t_{mn}(x, y) \; dy^m\otimes dy^n + 
A^{\overline a} \otimes \ga_{\overline a }
+ \ga_{\overline a }\otimes B^{\overline a}
\ee
where according to (5)
\be
t_{mn} (x,y) = \partial_m y'^p (y)\; \partial_n y'^q (y) \; t_{pq} 
(x',y')
\ee
The $\{ \ga_{\overline a} \}$ is a basis of $G_i$-invariant one-forms 
\be
\ga_{\overline a} = \ga_{\overline a ,m}(x, y)\; dy^m 
= \ga_{\overline a ,m}(x', y') \; dy'^m 
\ee 
and $\; A^{\overline a} = A^{\overline a}_\gm (x)\, dx^\gm \; ,  
B^{\overline a} = B^{\overline a}_\gm (x)\, dx^\gm$ are one-forms on $M$.

Going to cases of interest, if we consider a non degenerate $G_i$-invariant 
metric on $M\times\gS$, $t\equiv G$ symmetric, we can choose
\be
\ga_{\overline a ,m} (x, y) = G_{mn}(x, y)\; v_{\overline a}^n (y)
\ee
that clearly satisfy (8) and then (6) can be put on the form
\be
G = G_{\gm \gn}^{(d)}(x)\; dx^\gm \otimes dx^\gn + G_{mn}(x, y)\; 
\left( dy^m + v_{\overline a}^m (y) \; A^{\overline a} \right)
\otimes \left( dy^n + v_{\overline a}^n (y) \; A^{\overline a} \right)
\ee
where
\begin{eqnarray}
G_{\gm \gn}^{(d)}(x) &=& G_{\gm \gn}(x) - 
g_{\overline a\overline b}(x)\; A_\gm^{\overline a}(x) \; A_\gn^{\overline 
b}(x)\cr 
g_{\overline a\overline b}(x) &=& 
G_{mn}(x,y)\;v_{\overline a}^m (y) \; v_{\overline b}^n (y)
\end{eqnarray}
Due to $G_i$ invariance the scalar product 
$g_{\overline a\overline b}$ must be $y$-independent; if $d_g = N$ (the 
case we will be interested in) we can introduce the dual forms to the 
$\{v_{\overline a}\}$ 
\begin{eqnarray}
\gw^{\overline a} &=& dy^m \; \gw_m^{\overline a}(y)\cr 
\delta_{\overline a}^{\overline b}&=& 
v_{\overline a}^m (y) \; \gw_m^{\overline b}(y) 
\end{eqnarray}
Then (11) enforces $G_{mn}(x,y)$ to have the form
\be
G_{mn} (x,y) = g_{\overline a \overline b} (x) \; 
\gw_m^{\overline a}(y) \; \gw_n^{\overline b}(y)
\ee
and $G$ can be written as 
\be
G = G_{\gm \gn}^{(d)}(x)\; dx^\gm \otimes dx^\gn + g_{\overline a 
\overline b} (x) \; \left( \gw^{\overline a} + A^{\overline a} \right)
            \otimes \left( \gw^{\overline b} + A^{\overline b} \right) 
\ee
Before continuing we would like to make some remarks about two facts we 
find not properly clear in the literature [4]. 
The first one is that the Killing fields are the $\{ u_a \}$, {\cal not} the 
$\{ v_{\overline a} \}$ that appear in (10); the gauge fields are 
connections on a $G_g$ vector bundle, not on a $G_i$ one. 
The second fact is that in standard KK considerations 
\be
G_{mn}(x, y) \equiv e^{ 2\,\phi (x) } \; g_{mn}(y) 
\ee
for some $G_i$ invariant $g_{mn}$, that is equivalent to have 
\be
g_{\overline a\overline b} (x) = 
\eta_{\overline a\overline b} \; e^{2\, \phi(x)}  
\ee
for some constant $\eta_{\overline a\overline b}$, but in general there 
is no need for this to be so and we will remain with $\frac{N(N+1)}{2}$ 
scalars fields (moduli) on $M$.
We will see later (see equation (51)) an example in which (16) is not 
verified.

Finally we find in a similar way that a $G_i$-invariant antisymmetric 
tensor $B$ can be written as 
\be
B = \frac{1}{2} \; B_{\gm\gn}(x)\; dx^\gm\wedge dx^\gn
  + \frac{1}{2} \; b_{mn}(x, y)\; dy^m\wedge dy^n
  + b^{\overline a}\wedge \ga_{\overline a}
\ee
with $b_{mn}(x,y)$ obeying (7) and 
$\; b^{\overline a} = b_\gm^{\overline a} (x) \; dx^\gm\;$ 
a one form on $M$.

\bigskip\bigskip

\n {\bf 2. The model}
\bigskip

It is well-known that GWZWM's are two dimensional conformal field 
theories that explicitely realize the Goddard-Kent-Olive  $G/H$ coset 
construction [6] 
and give rise to a sigma model with specific backgrounds $(G , B ,D)$ 
corresponding to the metric, antisymmetric tensor and dilaton modes of 
the string.
The field equations for them are only perturbatively known; 
by working out the $d+N$ dimensional objects (curvature, stress 
tensor, etc) related to the fields in (10,17) we can obtain the 
dimensionally reduced effective action in terms of the 
$d$ dimensional fields $(G, A, B, b , D )$. 
This action contains the bosonic part of $d=10, N=1$ SUGRA 
coupled to SUSY Yang-Mills theory, the last coupling being correctly 
reproduced by the dimensional reduction, and then reproducing the bosonic 
sector of the low energy heterotic and type I strings.
  
We do not describe all this here because 
we will not need it and mainly because we will follow a purely algebraic 
route that yields to the exact fields (the two dimensional functional method 
gives only the one-loop result).
  
Models of this type based on the $SU(2,1)/SU(2)\times U(1) $ and 
$SU(2,1)/SU(2)$ cosets were considered in [8] and [3] respectively. 
Here we will consider the gauging of a $U(1)$ non maximal subgroup 
that leads to a $\; 8-1=7$ dimensional space-time, a $SU(2,1)/U(1)$ coset  
that in some sense is the natural generalization of the $SU(1,1)/U(1)$ 
Witten's black hole [9].
However as it was pointed out in [10], isometries should appear 
associated with the maximal group commuting with $H$, in our case 
$SU(2)$, that will allow to interpret the solution as a compactification to 
$d=4$ dimensions in the spirit of the KK dimensional reduction reviewed before.

From the conformal field theory point of view the model has a central charge 
\be 
c(k) = \frac{8\; k}{k-3} - 1 = 7\;\frac{k+ \frac{3}{7}}{k-3}
\ee
Asking as usual the cancellation against the ghost contribution 
$c_{ghost}=-26$ gives a conformal value $k_c = \frac{81}{19}$.
At difference of other coset models, there exists only one string theory 
corresponding to the perturbative phase of the model.
In any case, a positive central charge requires $k>3$; we will assume 
for reasons to be clarified in Section 4 that $k>4$, i.e. 
\be
\lambda \equiv \frac{4}{k} < 1
\ee

Let us move now to the computations of the fields recalling some facts 
described at length in References [3,8]. 
An arbitrary element $g \in SU(2,1)$ may be locally parametrized as follows  
\begin{eqnarray}
g &=& T(s\, \vec n)\; e^{i \frac{\sqrt{3}}{2}\varphi \gl_8 } \; H(Y,1)\cr
T(s\,\vec n ) &=& 
\left( \matrix{
(1 + s^2\, \vec n \, {\vec n}^\dagger)^\frac{1}{2} & s\,\vec n \cr
s\,{\vec n}^\dagger                               & c
} \right) \;\; , \;\; \vec n = \left( \matrix{
n^1 \cr
n^2
}\right) \cr
H(Y ,1) &=& \left( \matrix{      Y  &  \vec 0\cr
            {\vec 0}^t  &  1
} \right)\;\; ,\;\; Y\in SU(2)
\end{eqnarray}
The two dimensional complex vector $\vec n$ is unimodular what allows to 
define an associated $SU(2)$ matrix in the following way
\be
  N\equiv  \left( \matrix{     
{n^1}^* &{n^2}^* \cr
        - n^2 &  n^1
} \right) = 
e^{i\frac{\ga}{2}\sigma_3} \; e^{i\frac{\gb}{2}\sigma_2} \; 
e^{i\frac{\gamma}{2}\sigma_3} 
\ee
where we have introduce Euler angles $\; ( {\ga ,\gb ,\gamma} )$ and Pauli 
matrices $\{ \sigma_i \}$.

On the other hand, it is easy show that 
\be
g = e^{- i\gd\lambda_8 }\;T(s\,e^{i\sqrt{3}\,\gd } \vec n)\; 
e^{i \frac{\sqrt{3}}{2}\varphi \gl_8 } \; H(Y,1)\; e^{i\gd \lambda_8 }
\ee
for {\cal any} $\gd$, i.e. it does not depends on $\gd$. 
In terms of the matrix $N$,
\be
e^{i\sqrt{3}\gd}\; \vec n  \Leftrightarrow e^{-i\sqrt{3}\, \gd\sigma_3} \; N 
\ee

Therefore if we are to consider a theory gauge invariant under vector 
transformations
\be
g \rightarrow h\; g\; h^{-1}
\ee
with $h$ generated by $\lambda_8$, equations (20-23) tell us that the 
Euler angle $\ga$ in (21) should be gauged away. 
Then we will remain with the seven gauge invariant variables 
$0\leq \gb \leq \frac{\pi}{2}, \; 0\leq\gamma <2\pi$ 
parametrizing a two sphere $S^2$, together with 
$0\leq \varphi<2\pi , s\equiv \sinh r \geq 0 \; (c=\cosh r)$ 
and some $S^3$ variables parametrizing $Y$ that we will not need to explicit.
\bigskip\bigskip
 
\n{\bf 3. Currents and the exact solution}
\bigskip

Here we present the exact solution for the metric and dilaton fields guessed 
from the \"ansatz in References [11].
To this end we introduce some notation and collect some useful relations.
We refer the indices of the currents to the generators given by 
\{$( \vec\lambda )_i = \lambda_i , i=1, 2, 3 ;\;
\lambda_1^\pm = \frac{1}{2} (\lambda_4 \pm i \lambda_5 ) ;\; 
\lambda_2^\pm = \frac{1}{2} (\lambda_6 \pm i \lambda_7 ) ;\; \lambda_8 \}$,  
where $ \{ \lambda_a \}$ are the Gell-Mann matrices.
If $X = x_0 1 + i \;\vec x \cdot \vec\sigma$ is an arbitrary $SU(2)$ 
element $(\; x_0{}^2 = 1 - {\vec x}^2\; )$ the 
adjoint representation is given by the $3\times 3$ matrix 
\be
R(X)_{ij} \equiv \frac{1}{2}\; tr( \gs_i X\gs_j X^\dagger ) = (2 x_0{}^2 -1) 
\;\delta_{ij} + 2\; x_i\; x_j + 2\; x_0\; \ge_{ijk}\; x_k
\ee 
and the left and right $SU(2)$ (thought as linear operators in equations 
(27,28,30,33)) vector fields together with their dual forms are 
\footnote{
We assume henceforth, when no explicitely specified, to refer them to $Y$ 
- variables.
}
\begin{eqnarray}
\hat\xi_i^L &=& x_0 \;\partial_i - \ge_{ijk}\; x_j\;\partial_k = - 
\hat\xi_i^R |_{-\vec x} \cr
\gw_L^i &=& x_0 \; dx_i + \frac{x_i}{x^0}\; x_j \; dx_j - 
\epsilon_{ijk} \; x_j\;dx_k = - \gw_R^i |_{-\vec x}\cr
\delta_i^j &=& \hat\xi_i^L ( \gw_L^j ) = \hat\xi_i^R ( \gw_R^j ) 
\end{eqnarray}
They generate the left and right transformations
\begin{eqnarray}
\hat\xi_i^L (X) &=& i\; \sigma_i\; X \cr
\hat\xi_i^R (X) &=& i\; X\; \sigma_i 
\end{eqnarray}
and satisfy the commutation relations
\begin{eqnarray}
[\hat\xi_i^L , \hat\xi_j^L ] &=&  2\; \epsilon_{ijk}\; \hat\xi_k^L \cr 
[\hat\xi_i^R , \hat\xi_j^R ] &=& -2\; \epsilon_{ijk}\; \hat\xi_k^R \cr 
[\hat\xi_i^L , \hat\xi_j^R ] &=& 0
\end{eqnarray}

Now let us move to the computations. 
We define the left currents as linear operators on the group manifold $G$ by 
\be
\hat L _a g = - \lambda_a \; g \; ,\;\; g\in G
\ee
In the parametrization (20) the computations yield 
($ (\check e _i )_j = \delta_{ij}$ ) 
\vfill\eject
\begin{eqnarray}
\hat{\vec L} &=& i\; \left( \vec{\hat \xi^L} |_Y - \; \vec{\hat \xi^R} 
|_N \right)\cr
\hat L_\ga^+ &=& -\frac{1}{2}\; N_{1\ga}\; (\partial_r -i 
\;\frac{s}{c}\; \partial_\varphi ) + \vec A _\ga^+ \cdot \vec{\hat\xi^L} 
|_Y + \vec B _\ga^+ \cdot \vec{\hat\xi^L} |_N = ( \hat L_\ga^- )^*\cr
\hat L _8&=& i\;\frac{2}{\sqrt{3}}\left( \partial_\varphi - \frac{3}{2}\;
\hat\xi_3^L |_N \right) 
\end{eqnarray}
where 
\begin{eqnarray}
\vec A_\alpha^+ &=& \frac{i}{2sc}\; R(N)^t\, \left( 
N_{1\ga} \; \frac{s^2}{2} \;\check e _3 + N_{2\ga}\; c\, (c-1)\; 
( \check e _1 - i \;\check e _2 ) \right)\cr
\vec B_\alpha^+ &=& \frac{-i}{2sc}\left( N_{1\ga}\; ( 2 c^2 -1)\; 
\check e _3 +  N_{2\ga}\; c^2\; ( \check e _1 - i \;\check e _2 ) \right)
\end{eqnarray}
Similarly we define the right currents by
\be
\hat R_a g =  g \;\lambda_a \; , \;\;\; g\in G
\ee
and compute them to get ($u \equiv e^{ i \frac{3}{2} \varphi }$ )
\begin{eqnarray}
\hat{\vec R} &=& - i\; \vec{\hat \xi^R} |_Y \cr
\hat R_\ga^+ &=& \frac{u}{2}\; (N Y)_{1\ga}\; (\partial_r  + i\; 
\frac{s}{c}\; \partial_\varphi ) + \vec A _\ga^+ \cdot \vec{\hat\xi^L} |_Y + 
\vec B _\ga^+ \cdot \vec{\hat\xi^L} |_N = ( \hat R _\ga^- )^* \cr
\hat R _8 &=& - i \; \frac{2}{\sqrt{3}}\; \partial_\varphi
\end{eqnarray}
where
\begin{eqnarray}
\vec A_\alpha^+ &=& \frac{i\, u}{2sc}\; R(N)^t\, \left( 
(N Y)_{1\ga} \; \frac{s^2}{2} \;\check e _3 + (N Y)_{2\ga}\; 
c\, (c-1) \; ( \check e _1 - i \check e _2 ) \right) \cr
\vec B_\alpha^+ &=& \frac{i\, u }{2sc} \left( (N Y)_{1\ga}\; \check e _3 + 
(N Y)_{2\ga}\; c\; ( \check e _1 - i \check e _2 ) 
\right)
\end{eqnarray}
By construction both set of currents satisfy the corresponding 
$\lambda_a $-algebra.
Now we introduce the Casimir operators ($g_{ab} = tr \lambda_a \lambda_b $)
\be
\Delta_G^L = g^{ab} \hat L _a \hat L _b
\ee
and the Virasoro-Sugawara laplacian associated with the coset 
$G/H = SU(2,1)/U(1)$ 
\be
\hat L _0^L = \frac{1}{k-3}\; \Delta_G^L - \frac{1}{k}\; \Delta_H^L
\ee 
with analog construction in the right sector.

Finally we consider gauge invariant functions, i.e.
\be
(\hat L_8 + \hat R_8 ) f(g) = -i\, 2\, \partial_\ga f(g) = 0 
\;\; , \;\; g\in SU(2,1) 
\ee
which are the $\ga$ - independent ones as it should be as remarked above, 
and on this subspace we define the metric and dilaton fields to be those 
that obey the ``hamiltonian" equation
\footnote{
The computations in the left and right sectors lead to the same result.
}
\begin{eqnarray}
\hat H f(g) &\equiv& \frac{1}{k-3}\; \chi^{-1} \partial_\mu (\;\chi\; 
G^{\mu\nu}\; \partial_\nu ) f(g)\cr
\hat H &\equiv& \hat L_0^L + \hat L_0^R  = \frac{1}{k-3} \left( 
{\vec{\hat L}}^2 + 2 \{ \hat L_\ga^+ , \hat L_\ga^- \} + 
\frac{3}{4}\lambda \; L_8{}^2  \right)\cr
\chi &\equiv& e^D \; |\det\; G|^{\frac{1}{2}}
\end{eqnarray}

By carrying out the computations we read from these equations the exact 
backgrounds; it is useful to introduce the standard spherical versors on 
$S^2$
\begin{eqnarray} 
 \check r &=& \sin\gb \, \cos\gamma \; {\check e}_1 + 
\sin\gb \,\sin\gamma \;{\check e}_2 + \cos\gb \; {\check e}_3\cr
\check\gb &=& \cos\gb \, \cos\gamma \; {\check e}_1 + 
\cos\gb \,\sin\gamma \;{\check e}_2 - \sin\gb \; {\check e}_3\cr
\check \gamma &=& - \sin\gamma \;{\check e}_1 + \cos\gamma\; {\check e}_2
\end{eqnarray} 
as well as the $3\times 3$ matrix $Q$ and the function $a$,
\begin{eqnarray}
Q &=& 1 - \left( 1 - \left( 1 + \frac{\lambda}{4}\, a \right)^{-\frac{1}{2}}
\right)\; \check r \,{\check r}^t \cr
a(r)^{-1} &=& 1 - \lambda \,\frac{c^2}{s^2}  
\end{eqnarray}
Then a convenient ``seibenbein" is given by 
\begin{eqnarray}
\vec e  &=& Q^{-1} \; \vec{\hat \xi^L} \cr
   e _4 &=& \frac{1}{\sqrt{a}}\; \frac{s}{c} \; \left( \partial_\varphi + 
\frac{a}{2}\; {\check r}\cdot \vec{\hat \xi^L} \right)\cr
   e _5 &=& \partial_r \cr
   e _6 &=& \frac{2}{s}\; \left( \partial_\gb + 
\frac{c-1}{2} \; {\check\gamma}\cdot \vec{\hat\xi^L} \right) \cr
   e _7 &=& \frac{2}{s\, \sin\gb}\; \left( \partial_\gamma - 
\frac{c-1}{2} \sin\gb \; {\check\gb}\cdot \vec{\hat\xi^L} \right) 
\end{eqnarray}
whose dual basis is
\begin{eqnarray}
\vec\gw  &=& Q \; \left( \vec{\gw_L } + \vec A \right) \cr
   \gw^4 &=& \sqrt{a}\; \frac{c}{s}\; d\varphi\cr
   \gw^5 &=& dr \cr
   \gw^6 &=& \frac{s}{2}\; d\gb \cr
   \gw^7 &=& \frac{s}{2}\;\sin\gb\; d\gamma 
\end{eqnarray}
where the one-forms $\vec A$ are given below in equation (50).
Then the seven dimensional metric results 
\be
G = \eta^{ab} \; \gw_a \otimes \gw_b
\ee
where $\eta_{ab}$ is minkowskian with signature $(---++++)$ (the actual 
signature depends on $r$ and the value of $k$, however see next section). 
 
The dilaton field on the other hand is given by 
\be
D(r) = D_0 + \ln |\; s^2 \; \sqrt{ 1 + \frac{\lambda}{4}\; a^{-1} }\; |
\ee

We spend some words here about the antisymmetric tensor. 
The method does not permit to obtain its exact value (see however [14]), 
but it is possible to get the one loop result from the standard 
integrating out of the gauge fields mentioned at the beginning of Section 2.
We have carried out that calculation (verifying of course the  
$\lambda = 0$ limit of (43,44)) and for completeness we quote the result 
\be 
H^{(1l)} \equiv dB^{(1l)} =  d( \vec A^{(1l)} \wedge \vec{\gw}_L ) - 
\frac{1}{2} \; \sin\gb\; d\gb\wedge d\gamma\wedge d\varphi +
2\; \gw_L^1 \wedge \gw_L^2 \wedge \gw_L^3
\ee
By comparing it with (17) we identify  
\be
\vec b = \vec A \;\;\;  ,
\ee
equality that we conjecture to hold exactly.
\bigskip\bigskip

\n {\bf 4. The four dimensional interpretation}
\bigskip

From the parametrization (20) it follows that the right 
global transformation 
\be
g\rightarrow g\; H(X,1) \;\; \Leftrightarrow \;\;Y \rightarrow Y\; X 
\;\;\; , \;\;\; X\in SU(2)
\ee
should be an invariance of the model, in particular an isometry. 
And in fact this is manifest in that the $Y$ dependence of the backgrounds
comes through the {\cal right} invariant one forms $\gw_L^i$
\be
\gw_L^i = \frac{1}{2\, i}\; tr ( \sigma_i \; dY\, Y^{-1}) =   
          \frac{1}{2\, i}\; tr \left( \sigma_i \; d(Y X )\, (Y X )^{-1}\right)  
\ee 
Then in the language of Section 1, we identify $\; G_i \equiv SU(2)_{right}$
and according to (27, 28, 33) the generators of this transformation are the 
$u_{a}\sim {{\hat\xi}_i^R} $ vector fields. 
And therefore from (28) the identification of $G_g \equiv SU(2)_{left}$ 
and the $v_{\overline a}\sim {{\hat\xi}_i^L} $ {\cal right} invariant vector 
fields ( together with $\gw^{\overline a}\sim {\gw^i_L}$) is straightforward.

Taking into account these facts we can identify the four dimensional 
manifold $M \sim \Re \times S^1 \times S^2$ and the metric and 
(one loop) antisymmetric tensor as 
\vfill\eject
\begin{eqnarray}
  G^{(4)} &=& a\,\frac{c^2}{s^2}\; d^2\varphi + d^2 r + 
              \frac{s^2}{4}\; ( d^2 \gb + \sin^2 \gb \; d^2 \gamma )\cr
B^{(4)} &=& \frac{1}{2}\; \cos\gb\; d\gamma\wedge d\varphi
\end{eqnarray}
The dilaton is always given by (44) and the $SU(2)$ gauge fields are 
\be
\vec A = \vec b = -\frac{a}{2}\; \check r \; d\varphi  -
\frac{c-1}{2}\; \check\gamma\; d\gb + 
\frac{c-1}{2}\;\sin\gb\; \check\gb\; d\gamma
\ee 
On the other hand the fields $g_{mn}(x,y)$ and 
$b_{mn}(x,y)$ on $\gS \equiv S^3$ are read straight from (14, 17), in 
particular 
\be
g(x) = - Q(x)^2 = - 1 + 
\left( 1 + \frac{4}{\lambda}\; a^{-1} \right)^{-1}\;\check r \,{\check r}^t 
\ee 

The four dimensional backgrounds present a manifest and not at all obvious 
$SO(3)\times U(1)$ isometry.
Furthermore is not asymptotically flat but asymptotic to a constant  
curvature geometry, an usual feature in  bosonic string models due to 
the presence of the cosmological constant term [12, 3, 8]. 
This can be seen for example from the four dimensional Ricci scalar 
\be
\frac{1}{6} \; {\cal R}^{(4)} = 1 - \frac{a^2}{s^4}\; \left( 
\frac{s^2}{a^2} - \lambda \right)
\ee 
On the other the one loop solution presents a true singularity at $r=0\,$ 
that remains at higher orders. 
However at the value of the radius $\; r= r_0 > 0\;$ defined by 
\be
s_0 = \sinh r_0 = \left( \frac{\lambda}{1 - \lambda} \right)^{\frac{1}{2}}
\ee
the curvature also exploits and another true singularity of purely 
quantum origin appears; this means that if the quantum theory has a sense 
(fact not addressed here) we certainly cannot analytically continue 
the solution beyond $r_0$ and then the singularity becomes ``shifted".
Then the signature of the solution is also preserved..  
In ``Schwarzchild" like radial coordinate 
\be
0< R(r) = c_0\; \ln\left( \frac{s}{s_0} + \sqrt{ \frac{s^2}{s_0{}^2} - 1 } 
\right) < \infty 
\ee
it looks like an instanton in compactified euclidean time $\varphi$ with 
a singularity at the origin $R=0$

\eject

\begin{eqnarray}
  G^{(4)} &=& F(R)^{-2}\; d^2\varphi + F(R)^2 \; d^2 R + 
              \frac{s_0{}^2}{4}\; \cosh^2\frac{R}{c_0} \;\; ( d^2 \gb + 
\sin^2 \gb \; d^2 \gamma )\cr 
F(R)^2 &=& \lambda\, (1-\lambda)\, \frac{\sinh^2\frac{R}{c_0}}
{1 + \lambda\, \sinh^2\frac{R}{c_0} }
\end{eqnarray}
Furthermore, we can adscribe to it non abelian ``hair". 
If we introduce as usual the $SU(2)$ connection and its strenght by 
\begin{eqnarray}
A &=& i \; \vec \gs \cdot \vec A \cr
F &=&  dA + A\wedge A =  i \; \vec \gs \cdot \vec F   
\end{eqnarray}
then the non abelian magnetic charges are defined by
\be
Q \equiv - \frac{1}{4\, \pi}\; \int_{S^2} F  
\ee
where the two sphere is taken to be in the asymptotic region. 
The computation shows up a non zero charge associated to the ``radial" 
generator $\vec \gs \cdot \check r$
\be
Q =  i \; \frac{1}{2\, g_s{}^2} \; \vec \gs \cdot \check r 
\ee
where $g_s$ is the string coupling constant at infinity.

\bigskip\bigskip

\n{\bf 5. Conclusions}
\bigskip

We have obtained  a highly non trivial instanton of the (unknown) exact 
classical effective action of the bosonic string theory. 
As showed in [11] the one loop results are (up to a trivial rescaling) 
solutions of the type II superstring, and from the remarks in section 4 
they are qualitatively similar to the exact ones.
It has an essential singular submanifold and couples to $SU(2)$ 
Yang-Mills fields with non trivial charge. 
On the other hand an obvious as well as unexpected isometry $SO(3)\times 
U(1)$ is present in the four dimensional fields, equations (49). 
However this is not an isometry of the whole (seven dimensional) solution, 
as it is seen e.g. from the form of the gauge fields, equation (50).
\footnote{
Maybe it is worth to remark that the $SU(2)$ isometry of the model that 
allows the compactified interpretation has to do with the $Y$ variables 
on the compact space $S^3$, but nothing to do with the four dimensional 
fields that in general should not present isometries. 
}
This fact is even more manifest in the equivalent backgrounds related to 
(43-45) by $T$-duality [13]. 
For example the one loop 
\footnote{
We cannot compute the exact dual backgrounds because we do not know the 
exact antisymmetric tensor.
}
four dimensional metric reads
\be
\tilde G^{(4)} = \frac{ 16\, s^2 \, c^2}{ (1 + 3\, c^2 )^2} \;
( d\varphi - \frac{1}{2}\; \cos\gb\;d\gamma )^2 + d^2 r + 
\frac{s^2}{4}\; ( d^2 \gb + \sin^2 \gb \; d^2 \gamma ) 
\ee
Another feature of the one loop dual solution we believe interesting 
is that the dilaton field
\be
\tilde D = {\tilde D}_0 + \ln (1 + \frac{3}{4}\; s^2 )
\ee
goes to a constant (instead of being linear) in the asymptotic region $r\gg 
1$, as most asymptotically flat solutions does, but we do not know if this 
behaviour survives at higher orders.
    
\bigskip\bigskip

\noindent{\bf References}

\begin{enumerate}

\it C. Callan, D. Friedan, E. Martinec and M. Perry, \np 262 (1985), 593. 
\it G. Horowitz: ``The Dark Side of String Theory: Black Holes and Black 
Strings", Proceedings of the 1992 Trieste Spring School on String Theory 
and Quantum Gravity (World Scientific, Singapore, 1993); 
D. L\"ust: ``Cosmological String Backgrounds", CERN-TH-6850/93, March 1993 
(unpublished); 
F. Quevedo: `` Lectures on Superstring Phenomenology", CERN-TH/96-65, 
March 1996, hep-th/9603074.
\it A. Lugo: `` Exact monopole instantons and cosmological solutions in 
String Theory from abelian dimensional reduction", UNLP-TH-96/03, 
hep-th/9603182, to appear in \prd
\it ``Modern Kaluza-Klein theories", ed. by T. Appelquist, A. Chodos and 
P. Freund, Addison-Wesley (1987).
\it T. Eguchi, P. Gilkey and A. Hanson, Phys. Rep. 66 (1980), 213.
\it P. Goddard and D. Olive, Int. Jour. Mod. Phys. {\bf A} (1986), 303.
\it See for example, M. Green, J. Schwarz and E. Witten: ``Superstring 
theory", vol. 2, Cambridge University Press, Cambridge (1987), and 
references therein. 
\it A. Lugo, \prd 52 (1995), 2266.
\it E. Witten, \prd 44 (1991), 314.
\it P. Ginsparg and F. Quevedo, \np 385 (1992), 527.
\it I. Bars and K. Sfetsos, \prd 46 (1992), 4495; 
\prd 46 (1992), 4510; \pl 301 (1993), 183.
\it S. Giddings, J. Polchinski and A. Strominger, \prd 48 (1993), 5784.
\it A. Giveon, M. Porrati and E. Rabinovici, Phys. Rep. {\bf 244} (1994), 
77, and references therein.
\it I. Bars and K. Sfetsos, \prd 48 (1993), 844.

\ignore{
\it D. Karabali: ``Gauged WZW models and the coset construction of CFT", 
Brandeis report No. BRX TH-275, July 1989, and  references therein;\\
S. Chung and S. Tye, \prd 47 (1993), 4546.
\it T. Buscher, \pl 194 (1987), 59, \pl 201 (1988), 466.
\it M. Green and J. Schwarz, \pl 149 (1984), 117.
\it G. Chapline and N. Manton, \pl 120 (1983), 105.
\it G. Horowitz and A. Strominger, \np 360 (1991), 197.
\it S. Giddings, J. Polchinski and A. Strominger: ``Four 
dimensional black holes in string theory", UCSBTH-93-14, 
hep-th/9305083.
\it S. Giddings and A. Strominger, Phys. Rev. Lett. {\bf 67} 
(1991), 2930.
\it G. Gibbons and K. Maeda, \np 298 (1988), 741;\\ 
D. Garfinkle, G. Horowitz and A. Strominger, \prd 43 (1991), 3140.
\it J. Harvey, C. Callan and A. Strominger, \np 359 (1991), 611.
\it S. Thorne, R. Price and D. Macdonald: ``Black holes: the membrane 
paradigm", Yale University Press, New Haven (1986).
\it R. Wald: ``General Relativity", University of Chicago 
Press, Chicago (1984). 
\it K. Gawedzki, in ``New Symmetry Principles in Quantum Field Theory",  
Proceedings of the NATO Advanced Study Institute, Cargese, France (1991), 
ed. by J. Frolich et al., NATO ASI Series B: Physics Vol. 295 (Plenum, New 
York, 1992).
\it A. Sevrin: ``Gauging Wess-Zumino-Witten Models", VUB-TH.495 preprint, 
hep-th/9511050, and references therein.
\it A. Lugo, work in progress.
}
 
\end{enumerate} 

\end{document}